\documentclass[a4paper,11pt]{article}

\usepackage{graphicx}
\usepackage{natbib} 

\usepackage{amsmath}
\usepackage{amssymb}
\usepackage{enumerate}
\usepackage{verbatim}
\newcommand{\Jyvaskyla}{Jyv{\"a}skyl{\"a}}

\begin{document}

\title{Bayesian subcohort selection for longitudinal covariate measurements in follow-up studies}
\date{}

\renewcommand{\thefootnote}{\fnsymbol{footnote}}
\author{Jaakko Reinikainen$^{a,b}$\footnote{E-mail: jaakko.o.reinikainen@gmail.com} and Juha Karvanen$^a$ \\
$^a$ Department of Mathematics and Statistics,\\ University of \Jyvaskyla, \Jyvaskyla, Finland \\
$^b$ Department of Health, National Institute for Health and Welfare,\\ Helsinki, Finland}

\maketitle

\begin{abstract}
We consider planning longitudinal covariate measurements in follow-up studies where covariates are time-varying. We assume that the entire cohort cannot be selected for longitudinal measurements due to financial limitations and study how a subset of the cohort should be selected optimally in order to obtain precise estimates of covariate effects in a survival model. In our approach, the study will be designed sequentially utilizing the data collected in previous measurements of the individuals as prior information. We propose using a Bayesian optimality criterion in the subcohort selections, which is compared with simple random sampling using simulated and real follow-up data. This study extends previous results where optimal subcohort selection was studied with only one re-measurement and one covariate, to more realistic cases where several covariates and measurement points are allowed. Our results support the conclusion that the precision of the estimates can be clearly improved by optimal design.
\end{abstract}


%\footnotetext[1]{E-mail: jaakko.o.reinikainen@gmail.com}

\section{Introduction}

Longitudinal covariate measurements are often carried out in follow-up studies, when the covariates are time-varying. These measurements give useful information about the trajectories of the covariates. Frequent re-measurements provide more information, but in practice, limited resources may restrict measurement and researchers have to consider how to design the study cost-efficiently.

We study optimal design in a case where we cannot afford to re-measure the entire cohort but can only select a subset of the cohort, called a subcohort. The goal is to estimate the effects of the covariates on survival as precisely as possible. The study is designed sequentially, which here means that the subsets are selected just before the measurement times and all information collected prior to a new measurement is utilized. This kind of design procedure is realizable if reseachers can clearly define beforehand the purpose of data collection, i.e. the parameters of interest to be estimated from the data. In addition, the data must already be available during the follow-up. The proposed method requires especially that up-to-date survival information can be obtained when needed. This is possible, for example, in Finland, where data on mortality and hospitalizations are available from administrative registries.

Use of a Bayesian version of D$_s$-optimality, an optimality criterion based on Fisher information, is proposed for the selection of the subcohort. In addition to the D$_s$-criterion, there are also other optimality criteria, which were originally developed for design of experiments~\citep{Pukelsheim:optimalbook,atkinson2007optimum}, but can also be applied in observational studies. For example, \cite{karvanen2009optimal} considered optimal subset selection for genotyping in a follow-up study, \cite{buzoianu2009optimal} investigated selection of patients for a diagnostic test and \cite{mehtala2015optimal} studied optimal time spacings for observations of a multistate Markov process.

The problem considered in this article has previously been studied in a simple setting with only one time-varying covariate and one re-measurement after the baseline~\citep{reinikainen2014optimal}. The results indicated that the cost-efficiency of a follow-up study can be improved by applying optimal selection, which motivated further developments. The present paper extends the concepts to more realistic cases where several covariates and measurement points are allowed. 

Selecting only subcohorts for re-measurement creates a large amount of missing data, which were previously~\citep{reinikainen2014optimal} handled with multiple imputation and a likelihood-based approach with numerical integration. These approaches do not look promising for the generalized problem because multiple imputation of covariates conditioned on survival data would be complicated and numerical integration would become infeasible as increasing the number of covariates increases the dimension of the integral. Here, we handle missing data by using Bayesian data augmentation, which is expected to be a more flexible method when the number of measurement points and covariates is increased.

The optimal subcohort selection is studied using simulated and real data. The real data consist of the Finnish cohorts from the Seven Countries Study~\citep{keys1970coronary}, an international epidemiologic study characterized by a long follow-up time and several longitudinal covariate measurements. We use body mass index and smoking as time-varying covariates and all-cause mortality as the outcome. With these data, the proposed selection procedure is compared with simple random sampling of individuals to be re-measured and with a case where the entire cohort is selected for re-measurement.

\section{Survival model} \label{model}

In this section, we introduce the notation for our study design and survival model. These are later used to present our optimal subcohort selection procedure. Many of the following assumptions are made to be suitable for the real data example of Section~\ref{EW}, but the general idea is applicable to other designs and models as well.

Let us consider a follow-up study in which survival time is the response variable and $M$ longitudinal measurements are carried out for the time-varying covariates after the baseline measurement. We denote the covariate values by $x_{mjh}$ for the measurement $m=0,\dots,M$, the individual $j=1,\dots,N$ and the covariate $h=1,\dots,H$. The corresponding random variables are denoted by $X_{mjh}$ and for all the covariates and individuals shortly by $X_{m}.$ The measurements are carried out at time points $\tau_0,\dots,\tau_M$ in calendar time for the individuals who are alive and have been selected to be measured. The follow-up has a predetermined length ending at the time $\tau_{M+1}.$ 

The observed survival information at the time of $m$th re-measurement for the individual $j$ is denoted by $y_{mj} = (t_{mj}, \delta_{mj}),$ where $t_{mj}$ is the continuously measured age of an individual and $\delta_{mj}$ is the status indicator ($\delta_{mj}=1$ for an event and $\delta_{mj}=0$ for censoring). Although the survival time is observed continuously, the piece-wise modeling approach uses separate time and status variables for each measurement time interval. The individual $j$ has survival time variables $t_{1j},t_{2j},\dots$ for each part of the follow-up where they are still alive. The indicator $\delta_{mj}$ tells whether the individual has died between the measurement times $\tau_{m-1}$ and $\tau_m$. We denote the random variables related to survival information by $Y_{mj}$ and for all the individuals by $Y_m$.

We use notation $\boldsymbol{x}_{m}=(x_{m1},\dots,x_{mH})^T$ and $\boldsymbol{\beta}=(\beta_1,\dots,\beta_H)^T$ and continue by assuming that the covariates are related to the hazard of the event through the proportional hazards model
\begin{eqnarray}
\lambda(t_{m+1}|\boldsymbol{x}_{m}, \delta_{m}=0) & = & \lambda_0(t_{m+1}|\delta_{m}=0)\exp(\boldsymbol{\beta}^T \boldsymbol{x}_{m}). \label{timedepmodel}
\end{eqnarray}
Above, we have used a Markov assumption 
$$\lambda(t_{m+1}|\boldsymbol{x}_{0},\dots,\boldsymbol{x}_{m}, \delta_{m}=0) = \lambda(t_{m+1}|\boldsymbol{x}_{m}, \delta_{m}=0)$$
for the covariate effects. Conditioning on $\delta_{m}=0$ means that only those individuals contribute here who have not died before the $m$th re-measurement. The survival times are assumed to follow the Weibull distribution, when the baseline hazard function has the form
$$\lambda_0(t_{m+1}|\delta_{m}=0) = \frac ab\left(\frac{t_{m+1}}{b}\right)^{a-1},$$
where $a$ is the shape parameter and $b$ is the scale parameter. Then model~\eqref{timedepmodel} becomes a parametric form of the time-dependent Cox model~\citep{therneau2000modeling}. Other distributions than the Weibull could also be used for the survival times.

Now, we can write the survival function and the density function as
\begin{eqnarray*}
 S(t_{m+1}|\boldsymbol{x}_{m}, \delta_{m}=0) & = & S_0(t_{m+1}|\delta_{m}=0)^{\exp(\boldsymbol{\beta}^T \boldsymbol{x}_{m})} \ \mathrm{and} \\
 f(t_{m+1}|\boldsymbol{x}_{m}, \delta_{m}=0) & = & \lambda(t_{m+1}|\boldsymbol{x}_{m}, \delta_{m}=0)S(t_{m+1}|\boldsymbol{x}_{m}, \delta_{m}=0),
\end{eqnarray*}
where $S_0(t_{m+1}|\delta_{m}=0)$ is the baseline survival function. The modeling is carried out piecewisely in time, because the covariate information changes at the measurement points. For this reason we have to deal with left-truncated and possibly right-censored Weibull distributions in the time intervals $(\tau_{0},\tau_{1}],(\tau_{1},\tau_{2}],\dots.$ Survival times are left-truncated at the lower limit of a time interval because an observed survival time $t_{mj}$ cannot be smaller than $t_{m-1,j}.$ In addition, left-truncation is necessary because the age of an individual is used as the time scale and we do not assume that the follow-up would begin at time zero, which would be the time of birth. The likelihood contribution for the individual $j$ for the parameters $\beta_1,\dots,\beta_H,a$ and $b$ is
\begin{small}
\begin{equation} \label{likelihood}
L_j(\beta_1,\dots,\beta_H,a,b) = \prod_{m=0}^{m_j'} \left(\frac{f(t_{m+1,j}|\boldsymbol{x}_{mj},\delta_{m}=0)}{S(t_{mj}|\boldsymbol{x}_{mj},\delta_{m}=0)} \right)^{\delta_{m+1,j}} \left(\frac{S(t_{m+1,j}|\boldsymbol{x}_{mj},\delta_{m}=0)}{S(t_{mj}|\boldsymbol{x}_{mj},\delta_{m}=0)} \right)^{1 - \delta_{m+1,j}},
\end{equation}
\end{small}
where $m_j' = \max\{0,\dots,M : \delta_{mj}=0\}.$ 

\section{Optimal subcohort selection} \label{optimalselection}

If the entire cohort cannot be re-measured because of financial limitations, we have to select a subcohort, which we can afford to measure. The optimal selection aims to make the estimates of the parameters of interest as precise as possible subject to financial constraints. In this paper, we focus on the estimation of regression parameters $\boldsymbol{\beta}.$

Assume that baseline covariate measurements (and possibly some longitudinal measurements) have been carried out and that continuous survival information can be obtained during the study for all individuals. When we want to carry out the next longitudinal measurement for a subcohort, we proceed by taking the following general steps:
\begin{enumerate}
	\item Just before the new longitudinal measurement, use the data already collected to obtain prior information about the parameters of interest
	\item Define the optimality criterion as an expectation over the prior distribution 
	\item Maximize the optimality criterion and select an optimal subset of individuals for the re-measurement
	\item Re-measure the covariates for the selected subcohort
\end{enumerate}
The rest of this section provides a description of one possible way to perform steps 2. and 3. Section~\ref{estimation} focuses on step 1.

\subsection{Selection criterion in a general form} \label{selectiongeneral}

An optimal design problem can be seen as a problem of maximizing the expected utility $U(\boldsymbol{\xi})$ for a design $\boldsymbol{\xi}$ from a design space $\Xi$~\citep{Chaloner:Bayesianexperimental}. In our problem, $\boldsymbol{\xi}$ is an indicator matrix with individuals on rows and measurement times on columns, where an element $(j,m)$ is 1 if individual $j$ has been selected for measurement $m$ and 0 otherwise. The constraint of limited resources means here that the column sums are fixed in $\boldsymbol{\xi}.$ The column sums need not be the same, because we may have different amount of resources for different re-examinations.

Data $\boldsymbol{w} = (\boldsymbol{x},\boldsymbol{y})$, where $\boldsymbol{x}$ corresponds to covariate data and $\boldsymbol{y}$ to survival outcome, come from a sample space $\mathcal{W}.$ The outcome data $\boldsymbol{y}$ and the baseline covariate measurements are assumed to be available on all individuals, whereas longitudinal covariate data is collected according to the design $\boldsymbol{\xi}.$ The data are assumed to follow a model $p(\boldsymbol{w}|\boldsymbol{\theta}),$ where parameters $\boldsymbol{\theta}$ belong to the parameter space $\Theta.$

The fully Bayesian solution for the optimal design problem would involve integrating a measure of observed utility over data $\boldsymbol{w}$ and the posterior distribution of parameters $\boldsymbol{\theta}$. Instead of this, we use a common approach \citep{Chaloner:Bayesianexperimental,atkinson2007optimum} where the integration is done over the prior distribution of $\boldsymbol{\theta}$ and the utility is defined as a function of the expected information. This leads to the notation
\begin{equation} \label{utility_general}
U(\boldsymbol{\xi}) = \int_{\Theta} g \left[ E_{\boldsymbol{w}|\boldsymbol{\theta},\boldsymbol{\xi}} \{I_{\boldsymbol{w}}(\boldsymbol{\theta})\} \right] p(\boldsymbol{\theta}) d \boldsymbol{\theta},
\end{equation}
where $g$ is a function such as determinant (D-optimality).

Here, we use the D$_s$-criterion \citep{atkinson2007optimum}, which is a special case of the widely used D-optimality criterion.
The D-optimal design maximizes the determinant of the Fisher information matrix or equivalently minimizes the determinant of the covariance matrix. D$_s$-optimality considers only a subset of $s$ parameters. If the parameter vector $\boldsymbol{\theta} = (\theta_1,\dots,\theta_s,\dots,\theta_p)^T$ includes first the $s$ parameters of interest and then $p-s$ nuisance parameters, D$_s$-optimal design minimizes the determinant of the $s\times s$ upper left submatrix of $I(\boldsymbol{\theta})^{-1}.$ In our case, $s$ is the number of $\beta$-parameters in the survival model, and thus we will call the criterion D$_\beta$-optimality.

\subsection{Selection criterion for the Weibull proportional hazards model} \label{selectionweibull}

Let us consider the Fisher information matrix of the model introduced in Section~\ref{model} including parameters $\boldsymbol{\theta}^* = (\beta_1,\dots,\beta_H,a,b)$:
\begin{equation} \label{fisherinfo}
I_{X,Y}(\boldsymbol{\theta}^*) = -E \left(\frac{\partial^2 \log p(X_0,\dots,X_M,Y_1,\dots,Y_{M+1})}{\partial \boldsymbol{\theta}^{*2}}\right).
\end{equation}
The parameter vector $\boldsymbol{\theta}^*$ includes the parameters of the survival model but not the parameters of covariate processes. Using Markov assumptions $p(Y_{m+1}|X_0,\dots,\\X_m,Y_1,\dots,Y_m)=p(Y_{m+1}|X_m,Y_m)$ and $p(X_m|X_0,\dots,X_{m-1},Y_1,\dots,Y_m)=p(X_m|X_{m-1})$, we can factorize the logarithmic joint distribution as\\
\begin{align*}
& \log p(X_0,\dots,X_M,Y_1,\dots,Y_{M+1}) = \\
& \log p(X_0) +\log p(Y_1|X_0) + \dots +\log p(X_M|X_{M-1}) +\log p(Y_{M+1}|X_M, Y_M) ,
\end{align*}
which allows us to decompose $I_{X,Y}(\boldsymbol{\theta}^*)$ similarly.

When we are selecting individuals for the $m$th re-measurement, we have measured covariates $X_0,\dots,X_{m-1}$, which may include missing values and survival information is observed up to the time of the $m$th re-measurement, i.e. $Y_m$ is known. Therefore, the selection for the $m$th re-measurement is based on the expectations of $X_m$ and $Y_{m+1}$ utilizing the previously observed data. Only those individuals who have not yet died or been censored can be considered as candidates for the re-measurement. Due to Markov assumptions, it is sufficient that the selection is based only on the expectations with respect to the next unobserved part and not of all the forthcoming parts of the follow-up.

Now, using the above-mentioned factorization in matrix~\eqref{fisherinfo}, the information matrix used in the selection for the $m$th re-measu\-re\-ment can be written as
%\begin{small}
\begin{footnotesize}
\begin{eqnarray}  
I_{X,Y}^m(\boldsymbol{\theta}^*) & = & -E \left[\frac{\partial^2}{\partial\boldsymbol{\theta}^{*2}} \left\{\log p(X_0) + \log p(Y_1|X_0) +\dots+ \log p(X_{m-1}|X_{m-2}) + \log p(Y_m|X_{m-1},Y_{m-1}) \right\} \right. \nonumber\\
 &  & \left. + E\left\{\left(\frac{\partial^2}{\partial\boldsymbol{\theta}^{*2}}\log p(X_m|X_{m-1}) + \frac{\partial^2}{\partial\boldsymbol{\theta}^{*2}}\log p(Y_{m+1}|X_m, Y_m)\right)\middle| X_0,\dots,X_{m-1},Y_m\right\}\right]. \nonumber
\end{eqnarray}
\end{footnotesize}
%\end{small}
Above, the terms which do not include variable $Y$ vanish, because they do not include the survival model parameters $\boldsymbol{\theta}^*$ (parameters of the Weibull proportional hazards model). This leads to
\begin{small}
\begin{eqnarray}  
I_{X,Y}^m(\boldsymbol{\theta}^*) & = & E \left[-\frac{\partial^2}{\partial\boldsymbol{\theta}^{*2}} \left\{\log p(Y_1|X_0) +\dots+ \log p(Y_m|X_{m-1},Y_{m-1}) \right\} \right] \nonumber\\
 &  &  + E \left[E\left\{-\frac{\partial^2}{\partial\boldsymbol{\theta}^{*2}}\log p(Y_{m+1}|X_m, Y_m)\middle| X_0,\dots,X_{m-1},Y_m\right\}\right] \nonumber\\
 & = & I_{Y_1|X_0}(\boldsymbol{\theta}^*) +\dots+ I_{Y_m|X_{m-1},Y_{m-1}}(\boldsymbol{\theta}^*) + E\{I_{Y_{m+1}|X_m, Y_m}(\boldsymbol{\theta}^*)\},\label{info}
\end{eqnarray}
\end{small}
where the outer expectation of the last term is with respect to unobserved data $Y_{m+1}|X_m,Y_m.$

As values $Y_1,\ldots,Y_m$ and $X_0,\ldots,X_{m-1}$ are already observed, the first $m$ terms in~\eqref{info} are replaced by the observed information $J(\boldsymbol{\theta}^*)$. The information matrix is then a mixture of observed and expected information and its element in row $i$ and column $k$ is
\begin{small}
%\begin{footnotesize}
\begin{eqnarray}
 \Psi_{X,Y}^m(\boldsymbol{\theta}^*)_{i,k} & = & J_{Y_1|X_0}(\boldsymbol{\theta}^*)_{i,k} +\dots+ J_{Y_m|X_{m-1},Y_{m-1}}(\boldsymbol{\theta}^*)_{i,k} + E\{I_{Y_{m+1}|X_m, Y_m}(\boldsymbol{\theta}^*)\}_{i,k}  \nonumber\\
  & = & -\sum_{j=1}^N \left[\frac{\partial^2}{\partial\theta^*_i\partial\theta^*_k}\log p(y_{1j}|\boldsymbol{x}_{0j})\right]  \nonumber\\
  & & -\dots-\sum_{j=1}^{N_{m-1}} \left[\frac{\partial^2}{\partial\theta^*_i\partial\theta^*_k}\log p(y_{mj}|\boldsymbol{x}_{m-1,j},y_{m-1,j})\right]  \nonumber\\
  & & - \sum_{j=1}^{n_m}\left[E\left\{\frac{\partial^2}{\partial\theta^*_i\partial\theta^*_k}\log p(Y_{m+1,j}|X_{mj}, y_{mj})\middle| \boldsymbol{x}_{0,j},\dots,\boldsymbol{x}_{m-1,j},y_{mj}\right\}\right], \label{mixedinfo}
\end{eqnarray}
\end{small}
%\end{footnotesize}
where $N_{m-1}$ is the number of individuals who have not had an event before the measurement $m-1$ and $n_m$ is the number of individuals to be selected for the $m$th measurement. Note that when only a subset has been selected for the measurements at a time point $\tau_{m'}, m'\in \{1,\dots,m-1\},$ then covariates are missing for the individuals not selected. We describe in Section~\ref{estimation}, how these missing data are handled. The subcohort selection is carried out just before the new measurement, so $Y_{m+1}$ and $X_{m}$ are not observed for anyone. The expectation can be calculated by Monte Carlo integration.

The calculation of $\Psi_{X,Y}^m(\boldsymbol{\theta}^*)$ requires the second order partial derivatives of\\ $\log p(y_{1}|\boldsymbol{x}_{0}),\dots,\log p(y_{m+1}|\boldsymbol{x}_{m})$. The value of the D$_\beta$-criterion is obtained by taking the determinant of the $H\times H$ upper left submatrix of $\Psi_{X,Y}^m(\boldsymbol{\theta}^*)^{-1},$ where $\boldsymbol{\theta}^* = (\beta_1,\dots,\beta_H,a,b).$ We denote this value of the criterion by $D_\beta^m(\boldsymbol{\xi}^m,\boldsymbol{\theta}^*),$ where $\boldsymbol{\xi}^m$ is an indicator matrix describing which individuals have been measured at the time points $\tau_0,\dots,\tau_m$ and $\boldsymbol{\theta}^*$ emphasizes that the criterion depends on the parameters.

\subsection{Bayesian selection}

Now, we combine the Bayesian optimal design theory introduced in Section~\ref{selectiongeneral} and the criterion derived in Section~\ref{selectionweibull}. Fisher information matrices of nonlinear models usually depend on model parameters~\citep{Chaloner:Bayesianexperimental}, which is also the case in our application. Therefore, some prior information about the parameters is needed in order to use the D$_\beta$-criterion. In the Bayesian approach the information obtained from the data already collected during the follow-up and/or from previous studies can be used to provide prior distributions of the parameters when applying the optimality criterion. In other words, we use informative priors, which are actually posteriors from the data already collected, in the subcohort selections. In the selection for the $m$th re-measurement, the prior $p^m(\boldsymbol{\theta}^*)$ is, in fact, the posterior $p(\boldsymbol{\theta}^*|x_0,\dots,x_{m-1},y_1,\dots,y_m).$

To minimize the D$_\beta$-criterion, we specify the last column of $\boldsymbol{\xi}^m$ so that the expected utility
\begin{equation} \label{bayes_integral}
 U(\boldsymbol{\xi}^m) = -\int_{\boldsymbol{\theta}^*} D_\beta^m(\boldsymbol{\xi}^m,\boldsymbol{\theta}^*)p^m(\boldsymbol{\theta}^*)d\boldsymbol{\theta}^* 
\end{equation}
will be maximized. This is a specific form of the equation~\eqref{utility_general}. The integral above can be approximated by sampling parameter values from the multivariate prior distribution $p^m(\boldsymbol{\theta}^*)$, generating data $(X_m,Y_{m+1})$ given the parameters and then replacing the integral with a summation $\sum_{l=1}^{q}D_\beta^m(\boldsymbol{\xi},\boldsymbol{\theta}^*_l),$ where $q$ is the number of realizations sampled from $p^m(\boldsymbol{\theta}^*)$~\citep{atkinson1995optimum}. The priors become more informative during the follow-up as the amount of collected data increases.

In addition to the model parameters and the new data, the missing covariate values are also treated as unknown parameters. The predictive distributions of the missing values are used as the prior distributions in the selections. This means that if the previous measurements include missing data, the criterion~\eqref{bayes_integral} averages also over these informative prior distributions of the missing values. We draw $q$ realizations from the priors of the missing values and use them similarly to the realizations from $p^m(\boldsymbol{\theta}^*)$.

In practice, the number of different subsets that could be selected for the new measurement is easily so large that it is computationally impossible to go through each of them. Therefore some heuristic method is needed. We use a so-called greedy method~\citep{Wright:optimaltoxicology}, also known as sequential search~\citep{Dykstra:greedy}, to find an approximately optimal subset of individuals. This method selects $n$ individuals sequentially one by one: when $k-1$ individuals $(0<k<n)$ have been selected for the subcohort, the $k$th selection is made so that the optimality criterion is minimized on the condition that information from previously selected $k-1$ individuals is included in the calculation of the criterion. The procedure goes on similarly by selecting the next individual to be included in the subcohort so that the criterion is minimized taking into account the information obtained from the previously selected individuals.

From the beginning of the selection procedure, the information matrix~\eqref{mixedinfo} includes all the information already collected during the follow-up. All the individuals who have not had an event are considered as candidates for the new measurement. The procedure continues by testing which candidate should be included in the expectation part on the last row of~\eqref{mixedinfo} in order to obtain the minimum value of the Bayesian optimality criterion. If there are two or more individuals who would minimize the criterion, the selection between them can be done randomly. The expected information of the selected individual is then included permanently into~\eqref{mixedinfo} and the procedure continues until the subcohort has reached the predetermined size. The covariate measurements are carried out after the whole subcohort has been selected.

\section{Parameter estimation} \label{estimation}

The estimation of the parameters $\boldsymbol{\theta}^*$ of the survival model~\eqref{likelihood} is needed before each subcohort selection and finally when the follow-up study has ended. It should be emphasized that the model used in the subcohort selections need not be the same as the model used in the final analysis of the data, although the subcohort will be optimal with respect to the selection model. Once the data have been collected, the validity of the model assumptions can be reassessed and the model can be changed if needed. If the variables are the same in both models, the selected subcohort is likely to be better than a simple random sample, although the final analysis model could be more complex than the selection model. This is exemplified in Section~\ref{EW}. 

In the subcohort selections, the estimated posterior distributions of the parameters are used as informative prior distributions. The parameters are estimated using a Bayesian approach with Markov chain Monte Carlo (MCMC) sampling. We use non-informative priors for the parameters: $\boldsymbol{\beta}\sim N(\boldsymbol{0}, 10^4\cdot \boldsymbol{I}),\: r \sim \mathrm{Gamma}(1, 0.0001)$ and $\alpha \sim N(0, 10^4),$ where $r$ and $\alpha$ are reparameterized Weibull parameters, so that $r=a$ and $\alpha=-a\log b,$ $\boldsymbol{0}$ is a zero vector and $\boldsymbol{I}$ is an $H\times H$ identity matrix.

The subcohort selection may lead to a large amount of data `missing by design', which we handle by Bayesian data augmentation~\citep{tanner1987calculation}. The data missing by design are missing at random because the selection can depend only on variables that are already measured. Bayesian data augmentation has previously been used for data missing by design by e.g. \cite{kulathinal:2006}.

We consider continuous and binary covariates. The covariate models are estimated as a part of the entire Bayes model. If $x_1$ is a continuous covariate, the $m$th re-measurement of the covariate $x_1$ of individual $j$ is modeled with a linear regression
$$ x_{mj1} = c + \gamma x_{m-1,j,1} + \varepsilon_j, $$
where $c$ is a constant and $\varepsilon_j \sim N(0, v).$ The constant $c$, the coefficient $\gamma$ and the error variance $v$ could also be different at each measurement time if the structure of the covariate process should be allowed to vary over time. We assign priors $c\sim N(0,100), \gamma\sim N(0,100)$ and $v\sim \mathrm{Gamma}(1, 0.01),$ which are uninformative with respect to the scales of the covariates used.

If $x_2$ is a binary covariate, we use a logistic regression model
\begin{eqnarray*}
x_{mj2} & \sim & \mathrm{Bernoulli}(\pi) \\
\mathrm{logit}(\pi) & = & d_0 + d_1x_{m-1,j,2}.
\end{eqnarray*}
Above all the parameters remain constant in time, but this assumption could be relaxed if required. Uninformative priors $(d_0,d_1)^T\sim N(\boldsymbol{0}, 10^4\cdot \boldsymbol{I})$ are used. When applying these missing data models with MCMC estimation, it is important to center the covariates, because serious convergence issues are likely to arise if uncentered covariates are used~\citep{lunn2012bugs}.

\section{Simulation study}
\subsection{Description of the simulation study}

Simulation studies were performed to illustrate what kind of subcohort should be selected according to the D$_\beta$-criterion and what is the benefit of using it. We compared the use of D$_\beta$-optimal subcohort selection with simple random sampling (SRS) and evaluated how much precision is lost in the estimation when compared with measuring the entire cohort.

We considered a setting with two independent continuous covariates and a 30-year follow-up with three measurement times at time points 0, 10 and 20 years. The size of the cohort was 1500 individuals and the ages were generated from the uniform distribution with the range from 45 to 65 years at the baseline. Baseline measurements of continuous covariates $x_0$ and $z_0$ were made to follow $N(0,1)$ distribution, first re-measurements $x_1$ and $z_1$ were drawn from $N(\gamma x_0, \sigma^2_\varepsilon)$ and $N(\gamma z_0, \sigma^2_\varepsilon)$ and second re-measurements $x_2$ and $z_2$ from $N(\gamma x_1, \sigma^2_\varepsilon)$ and $N(\gamma z_1, \sigma^2_\varepsilon)$, where $\gamma=0.5$ and $\sigma^2_\varepsilon=0.75.$ These parameter values lead to serial correlation of 0.5 between consecutive measurements and to a constant variance of the covariates at each measurement. The covariates were generated independently of each other and independently of age.

Survival times of the individuals were simulated from the Weibull distribution conditioned on the covariates through a time-dependent Weibull proportional hazards model. The shape parameter of the Weibull distribution was set to $a=6.3$ and the scale parameter to $b = 27 900$ (in days), which roughly equal the parameters estimated from the real data used in Section~\ref{EW}. To investigate if the magnitude of hazard ratios of the covariates has some effect on the selection, we used different regression coefficients for the two covariates. The coefficients were $\beta_x=0.1\ (e^{\beta_x}=1.11)$ and $\beta_z=0.4\ (e^{\beta_z}=1.49),$ respectively for covariates $x$ and $z$. If the event had not occurred at the end of the follow-up (30 years after the baseline), the survival time was censored. The measurement times were the same for all the individuals in calendar time, but as the individuals were of different ages, the measurements were not carried out at the same time points in age.

The simulation was repeated 100 times on a supercomputer of CSC -- IT Center for Science Ltd. The three design approaches (D$_\beta$, SRS and full cohort) were applied to each simulated data set. In the beginning of the follow-up, the cohort included 1500 individuals, at the time of the second measurement, on average, 1173 individuals were alive, at the time of the third measurement, on average, 712 individuals were alive and at the end of the follow-up, on average, 299 individuals were alive. Model parameters were estimated using the OpenBUGS version 3.2.3~\citep{lunn2009bugs} and the rest of the calculations were carried out using the R statistical software version 3.1.1~\citep{Rproject2014}.

\subsection{Subcohort selection}

\begin{figure}%[h!]
	\centering
		\includegraphics[angle=-90,scale=1]{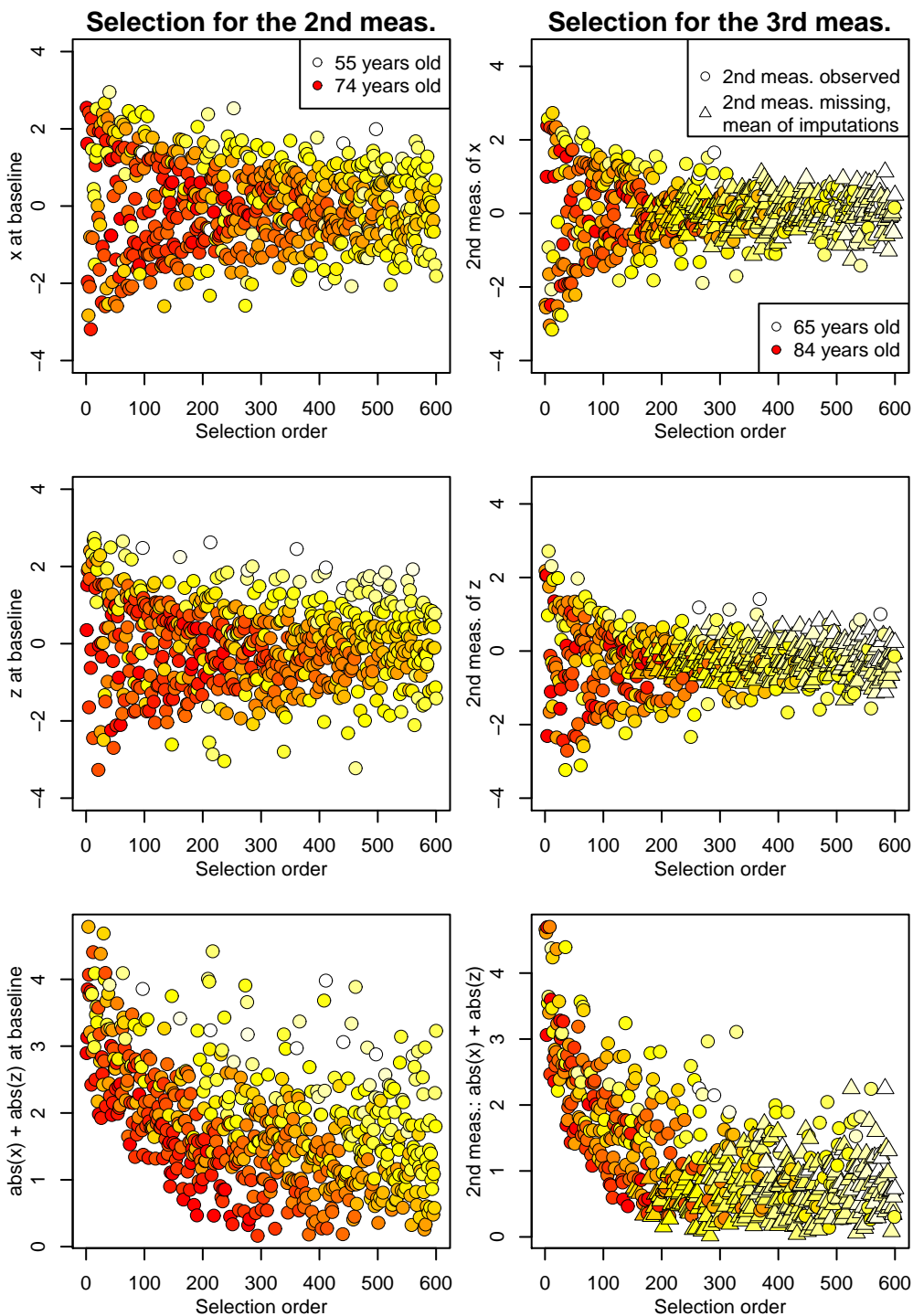}  
	\caption{Selection orders of individuals for the second and third measurements for any $n$ up to 600 using D$_\beta$-optimality and simulated data. Each point corresponds to one individual: the color shows the age of the individual at the time of the selection, the vertical axis shows the value of the covariate in the previous measurement ($x$ in the uppermost, $z$ in the middle and $|x|+|z|$ in the lowest panel) and the horizontal axis shows the round when the individual was selected in the greedy algorithm.} \label{sim2nd3rdmeas}
\end{figure}

The subcohort selections are carried out using the Weibull proportional hazards model as the underlying model with age as the time scale. The left panels of Figure~\ref{sim2nd3rdmeas} show what kind of individuals are selected by the D$_\beta$-criterion for the second measurement. The selections for the second and third measurements are carried out up to 600 individuals sequentially one by one, but it is worth noting that the order is irrelevant when analyzing the data. The older individuals are clearly preferred in the selection, which may arise from the fact that older individuals are more likely to have an event during the next part of the follow-up and therefore provide more information than those who are likely to be censored. On the other hand, individuals with extreme covariate values are selected first. This is a reflection of the result that extreme selection is optimal for first-order linear regression models~\citep{Elfving:optimumallocation}.

The individuals are plotted separately by baseline covariates $x$ and $z$, but there seems not to be a clear difference in the selection patterns, despite the different regression coefficients. The bottom left panel of Figure~\ref{sim2nd3rdmeas} shows the selection order against the sum of the absolute values, $|x|+|z|$. This is presented to illustrate better how extremity of the combination of the two independent covariates are preferred.

The selection of individuals for the third measurement, seen in the right panels of Figure~\ref{sim2nd3rdmeas}, seems to be quite similar to the previous selection. The individuals first selected for the third measurement have also been selected for the second measurement. After approximately 200 individuals, the selection also includes individuals who were not selected for the second measurement. In the figure, the values plotted for missing measurements represent averages of 100 independent values generated for each missing value in an MCMC estimation. The preference for higher age in the selection is also seen with these individuals.

\subsection{Design comparisons}

The analysis was carried out using the SRS-designs, D$_\beta$-designs and the entire simulated data set without subcohort selections. The sizes of the subcohorts varied from 300 to 600. In each design, the entire cohort was measured at baseline and the same subcohort size was used in both the second and the third measurements. We compared the bias and standard errors of $\beta_x$ and $\beta_z$ between the D$_\beta$-design and SRS.

The results in Table~\ref{simresults} show that there is no considerable bias in the estimates when only a subcohort is selected for re-measurements. According to standard deviations of the estimates and mean standard errors, the D$_\beta$-design seems to lead almost consistently to more precise estimation of the coefficients than the SRS-design. The difference becomes more prominent when the subcohort size decreases. We achieve virtually the same precision when comparing the D$_\beta$-design with subcohort size 400 to the SRS-design with subcohort size 500, or when comparing the D$_\beta$-design with subcohort size 300 to the SRS-design with subcohort size 400.

An important observation is that the precision does not decrease dramatically although only 300 individuals are re-measured. 300 is only on average 26\% of the individuals alive at the time of the second measurement and on average 42\% of the individuals alive at the time of the third measurement.

\begin{table}%[!h]
%\begin{footnotesize}
\caption{Simulation results for different designs from 100 simulation runs and for different sizes of the subcohorts ($n$). The entire cohort was measured at baseline. $\bar{\beta}_x$ and $\bar{\beta}_z$ indicate the means of the posterior means of $\beta_x$ and $\beta_z$. SD is the standard deviation of the posterior means and Mean(SE) is the mean of the standard errors estimated from the MCMC chains.} \label{simresults}
\centering 
%\begin{centering} 
%\centerfloat
%\begin{small}
\begin{tabular}{llcccccccccc}
\hline
 & & \multicolumn{3}{c}{Covariate $x\ (\beta_x=0.1)$} & & \multicolumn{3}{c}{Covariate $z\ (\beta_z=0.4)$}  \\
\cline{3-5}\cline{7-9} \\[-10pt]
 & Design & $\bar{\beta}_x$ & SD &  Mean(SE) & & $\bar{\beta}_z$ & SD &  Mean(SE) \\
\hline
 & Full cohort &  0.096 &  0.031  & 0.029 && 0.40 &    0.028  & 0.030 \\[5pt]
%\hline
$n=600$ & SRS & 0.096 &  0.036  & 0.032 && 0.40 &  0.031   & 0.033  \\
 & D$_\beta$ & 0.095 &  0.033  & 0.031 && 0.40 &  0.032   & 0.032  \\[5pt]
%\hline
$n=500$ & SRS & 0.096 &  0.036  & 0.033 && 0.40 &  0.033   & 0.035  \\
 & D$_\beta$ & 0.095 &  0.034  & 0.032 && 0.40 &  0.032   & 0.034   \\[5pt]
%\hline
$n=400$ & SRS & 0.098 & 0.041  & 0.036 && 0.40 &  0.042   & 0.039   \\
 & D$_\beta$ & 0.094 &  0.036  & 0.034 && 0.40 &  0.034   & 0.035  \\[5pt]
%\hline
$n=300$ & SRS & 0.093 &  0.043  & 0.043 && 0.42 &  0.063   & 0.059  \\
 & D$_\beta$ & 0.095 &  0.037  & 0.036 && 0.40 &  0.040   & 0.038  \\
\hline
\end{tabular}
%\end{small}
%\end{centering}
%\end{footnotesize}
\end{table}

\section{Results for the East-West study} \label{EW}

Next, we will present an application to data from a real follow-up study, the East--West study~\citep{reinikainen2015lifetime}. The East--West study was started as the Finnish part of the international Seven Countries Study~\citep{keys1970coronary} initiated in the late 1950s to investigate cardiovascular diseases and their risk factors across different countries and cultures. The Finnish cohorts consist of all men born between 1900 and 1919 and living in two geographically defined areas in Eastern Finland and in South--Western Finland ($N=1711$). The data include baseline measurements carried out in 1959 and longitudinal measurements in 1964, 1969, 1974, 1984, 1989, 1994, 1999 and follow-up for mortality until the end of 2011.

Our analysis with the East--West data is an example of using a binary covariate and a continuous covariate with nonlinear effect. For this example only a part of the data is used. We consider the measurement in the year 1964 as the baseline measurement, measurements in 1974 and 1984 as re-measurements and 1994 as the end of the follow-up, when censoring is carried out. After removing individuals who died before 1964, we had 1594 individuals. In the setting of this example, at the time of the second measurement 1225 individuals were alive, at the time of the third measurement 766 individuals were alive and at the end of the follow-up 320 individuals were alive.

All-cause mortality is the outcome in the analyses and age is chosen as the time scale. Smoking status is used as a binary covariate and body mass index (BMI) as a continuous covariate, whose effect on survival is assumed to be quadratic. Some studies have reported U-shaped associations between BMI and all-cause mortality~\citep{zhao2014body, corrada2006association} and this was also found in the East--West cohorts. The quadratic effect is implemented in the survival model simply by adding a squared term in BMI. Note that the final analysis of the data could still be carried out by using other methods, e.g. splines~\citep{therneau2000modeling}, even if polynomials were used in subcohort selection. Before the second measurement, there was no information in the data about the changes in smoking, so we assumed in the calculation of the expectation in~\eqref{mixedinfo} that a smoker at the baseline will be a non-smoker in second measurement with probability 0.4 and a non-smoker will become a smoker with probability 0.1. In the selection for the third measurement, these probabilities were estimated from the data.

In order to improve the convergence in MCMC estimation, all the covariates were centered before analysis by subtracting the means of observed baseline values from the corresponding covariate values. The mean of baseline BMI observations was 24.25 and the centered baseline BMI ranged from $-$9.25 to 21.62. The estimates of the analysis with the full cohort (Table~\ref{EWresults}) correspond to an upward opening parabola, which has minimum risk with centered BMI of 3.31.

\begin{figure}%[h!]
	\centering
		\includegraphics[angle=-90,scale=0.6]{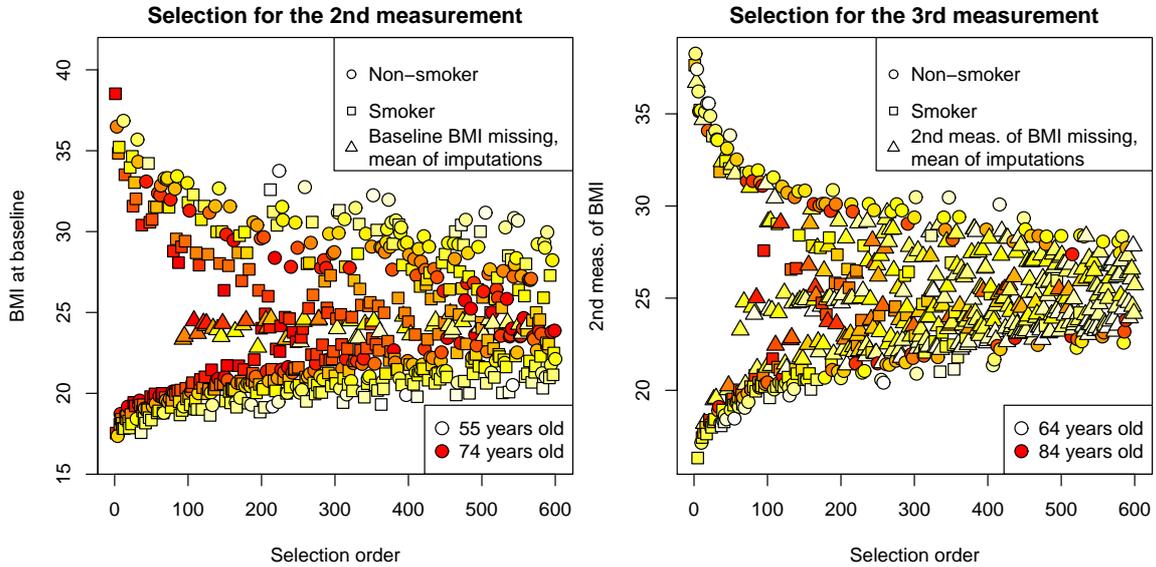}  
	\caption{ Selection order of individuals for the second measurement (left panel) and for the third measurement (right panel) for any $n$ up to 600 using D$_\beta$-optimality and the East--West data. Each point corresponds to one individual: the color shows the age of the individual at the time of the selection, the vertical axis shows the value of BMI in the previous measurement and the horizontal axis shows the round when the individual was selected in the greedy algorithm. Individuals with missing previous BMI are indicated by triangle symbols regardless of the missingness of the smoking status. Usually, the missingness of BMI means also the missingness of the smoking status in the data.} \label{ITLA2nd3rdmeas}
\end{figure}

The selection order of 600 individuals for the second measurement according to D$_\beta$ can be seen in the left panel of Figure~\ref{ITLA2nd3rdmeas}. The preference for extreme BMI values is clear. Although baseline measurements are missing for some individuals in the original data, some of these are still selected into the subcohort. At a fixed BMI level, older individuals seem to be selected before younger. We have a quadratic model in BMI, so one could have expected that in addition to extreme values also average values of BMI would have been preferred. The average values are, however, not so important in the second measurement, because we have baseline measurements for all individuals and thus have observed average values already much more than extreme values. The selected subcohort includes 327 individuals who were smokers at the baseline and 232 baseline non-smokers. At the baseline, there were also clearly more smokers than non-smokers. Apparently the selection procedure tries to balance the expected number of smokers and non-smokers measured altogether.

\begin{figure}%[h!]
	\centering
		\includegraphics[angle=-90,scale=0.85]{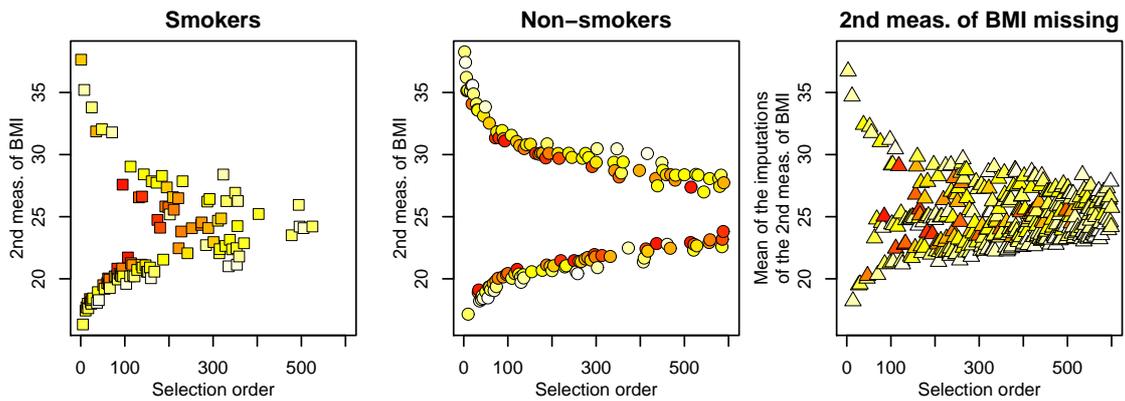}  
	\caption{Selection order of individuals for the third measurement using D$_\beta$-optimality and the East--West data. These panels represent the right panel of Figure~\ref{ITLA2nd3rdmeas} decomposed into smokers, non-smokers and those who have missing value of BMI in the second measurement. Individuals with missing BMI are indicated by triangle symbols regardless of the missingness of the smoking status. Usually, the missingness of BMI means also the missingness of the smoking status in the data.} \label{ITLA3rdmeasdecomp}
\end{figure}

Figure~\ref{ITLA2nd3rdmeas} (right panel) shows the subcohort selection for the third measurement. Similar patterns can be observed here as in the previous selection. One big difference is the large number of individuals who have missing previous measurements. This can be explained by the fact that many old individuals who were measured in the second measurement are already dead at the time of the third measurement. The effect of age does not seem to be so strong, but this comes mainly from the mixing of smokers and non-smokers in the plot. These are separated in Figure~\ref{ITLA3rdmeasdecomp}, which reveals that there are less smokers than non-smokers in the selected subcohort. In fact, there were 104 smokers, 203 non-smokers and 459 individuals with missing smoking status among the candidates for the third measurement, of which all 104 smokers, 138 non-smokers and 358 individuals with missing smoking status were selected for the third measurement. 

Table~\ref{EWresults} shows that there is clear benefit of using the D$_\beta$-design instead of the SRS. All the standard errors in the D$_\beta$-design are smaller than in the SRS-design for each subcohort size used. The D$_\beta$-selection leads usually to estimates closer to those obtained using the full cohort, than the SRS. Surprisingly, both selection methods seem to lead to greater estimates of the effect of smoking than the full cohort.

The standard errors of the quadratic term of BMI and smoking, obtained using SRS, are smaller when $n=300$ than when $n=400$, which is an unexpected result. 300 individuals is only 24\% of the individuals alive at the time of the second measurement and 39\% of those alive at the time of the third measurement, which may be too small proportions in a real study to obtain reliable estimates. Estimation may become sensitive to model misspecification when the proportion of missing data becomes large~\citep{saarela2012secondary}.

\begin{table}%[!h]
%\begin{footnotesize}
\caption{Results for the East--West data for different sizes of the subcohorts ($n$). The entire cohort was measured at baseline. For simple random sampling (SRS), $\bar{\beta}_1$, $\bar{\beta}_2$, $\bar{\beta}_3$ and SE are means of the posterior means and standard errors estimated from the MCMC chains from 1000 analyses.} \label{EWresults}
\centering
%\begin{centering} 
%\begin{small}
\begin{tabular}{llccccc}
\hline
 & & BMI (linear) & & BMI (quadratic) & & Smoking \\
\cline{3-3}\cline{5-5}\cline{7-7} \\[-10pt]
 & Design & $\bar{\beta}_1$ (SE) && $\bar{\beta}_2$ (SE) && $\bar{\beta}_3$ (SE)  \\
\hline
 & Full cohort &  $-$0.043 (0.0086) && 0.0065 (0.0011) && 0.39 (0.069)  \\[5pt]
%\hline
$n=600$ & SRS & $-$0.045 (0.0138) && 0.0071 (0.0017) && 0.45 (0.075)  \\
 & D$_\beta$ & $-$0.041 (0.0087) && 0.0063 (0.0011) && 0.45 (0.067)  \\[5pt]
%\hline
$n=500$ & SRS & $-$0.049 (0.0176) && 0.0078 (0.0022) && 0.47 (0.080) \\
 & D$_\beta$ &  $-$0.043 (0.0091) && 0.0062 (0.0011) && 0.42 (0.071)  \\[5pt]
%\hline
$n=400$ & SRS &  $-$0.052 (0.0218) && 0.0086 (0.0022) && 0.48 (0.083)  \\
 & D$_\beta$ & $-$0.042 (0.0093) && 0.0071 (0.0012) && 0.45 (0.073)  \\[5pt]
%\hline
$n=300$ & SRS & $-$0.044 (0.0232) && 0.0094 (0.0017) && 0.47 (0.080)  \\
 & D$_\beta$ &  $-$0.040 (0.0110) && 0.0058 (0.0015) && 0.48 (0.075)  \\
\hline
\end{tabular}
%\end{small}
%\end{centering}
%\end{footnotesize}
\end{table}

In practice, an analyst would not necessarily like to use the same model in optimal selection and in the final analysis. Table~\ref{modelcomparisons} shows the results from an example of using different models in the selections and final analysis. Here, we use the same data as in the previous example, but (centered) BMI as the only covariate. When the analysis model is quadratic, the quadratic selection model leads to slightly better precision than the linear selection model but both selection models still outperform SRS. When the analysis model is linear, the results do not deteriorate even if the quadratic selection model is used.

\begin{table}%[!h]
%\begin{footnotesize}
\caption{Results for comparisons of using different models in the subcohort selections and in the final analysis using the East--West data. 500 individuals were selected for the second and third measurements. For simple random sampling (SRS), $\bar{\beta}_1$, $\bar{\beta}_2$ and SE are means of the posterior means and standard errors estimated from the MCMC chains from 1000 analyses.} \label{modelcomparisons}
\centering
%\begin{centering} 
%\begin{small}
\begin{tabular}{llccc}
\hline
 & & BMI (linear) & & BMI (quadratic)  \\
\cline{3-3}\cline{5-5} \\[-10pt]
Analysis model & Selection & $\bar{\beta}_1$ (SE) && $\bar{\beta}_2$ (SE)   \\
\hline
Quadratic & Full cohort &  $-$0.059 (0.0082) && 0.0073 (0.0011) \\ 
 & SRS & $-$0.059 (0.0096) && 0.0079 (0.0013)   \\
 & D$_\beta$ (quadr. model) & $-$0.059 (0.0085) && 0.0077 (0.0011)   \\
 & D$_\beta$ (lin. model) & $-$0.059 (0.0087) && 0.0073 (0.0011)   \\[5pt]
%\hline
Linear & Full cohort &  $-$0.038 (0.0077) &&  \\
 & SRS & $-$0.033 (0.0084) &&  \\
 & D$_\beta$ (quadr. model) & $-$0.038 (0.0084) &&    \\
 & D$_\beta$ (lin. model) & $-$0.037 (0.0084) &&    \\
\hline
\end{tabular}
%\end{small}
%\end{centering}
%\end{footnotesize}
\end{table}

\section{Discussion} \label{discussion}

The cost-efficiency of a follow-up study can be improved by careful planning of longitudinal measurements. The present paper considered the case where we can afford measuring the time-varying covariates only for a subset of the cohort. The use of a Bayesian approach in optimal subcohort selection with a Fisher information based D$_\beta$-optimality was proposed. Our work also generalizes the results presented in \cite{reinikainen2014optimal}, where a simple case with only one covariate and one re-measurement was considered.

The estimates and their precision corresponding to the D$_\beta$-selection and simple random sampling (SRS) were compared. The use of the D$_\beta$-optimality led to more precise estimates and the precision was seen to remain satisfactory compared with the full cohort design.

The results indicated that in order to obtain estimates as precise as possible for regression parameters of the survival model, old individuals with extreme covariate values should be preferred, which is consistent with our previous results~\citep{reinikainen2014optimal}. A similar result was obtained when we used a covariate with quadratic effect (BMI) in the real follow-up data example. Another covariate used in this example was smoking status as a binary variable. Optimal selection seemed to balance the expected number of smokers and non-smokers measured.

The general idea of measuring only a subcohort is applied in many epidemiological study designs, like case-control and case-cohort designs and their variants~\citep{morgamcasecohort,sun2010design,keogh2013using}. However, in our application the setup is different. We use an approach with an explicit utility function describing the goal of the study. The presented approach borrows elements from optimal design of experiments and applies them to the design of an observational study.

We recommend the use of a Bayesian approach in this kind of sequential study design problem. Some prior knowledge is always required in nonlinear design problems, because optimal designs depend on model parameters and so a Bayesian approach with informative priors is a natural way to incorporate this knowledge into design optimization. Bayesian data augmentation appears here to be a flexible method for the handling of missing data when we increase the number of measurement points and covariates. 

In experiments, the optimal design for a linear model might consist of only two design points, which would give no power for detecting nonlinear effects. In principle, the same applies also in our setup, but in practice, the problem is not realized in observational studies with continuous covariates and moderate sample size. The reason for this is that only a few individuals with the optimal covariate values are available in the cohort and after they are selected, the selection procedure must rely on individuals with a wider variety of covariate values.

Although the covariate processes were assumed to be piece-wise constant in the subcohort selections, a joint model of survival and longitudinal data \citep{rizopoulos2012joint} could be considered in the final analysis for more realistic treatment of time-varying covariates. In the general case with multiple longitudinal covariates, joint modeling is, however, computationally very demanding and would be further complicated in our approach with a large amount of missing data.

We considered the design optimization with respect to only one model at a time. However, if two or more models or utility functions would be of interest, compound design criteria could be applied~\citep{atkinson2007optimum}. Then, the optimality criterion should include the parameters of all models of interest. If we later want to use the collected data to some purpose not addressed in subcohort selection, the optimality does not hold anymore, and in an extreme case the selected subcohort could perform even worse than SRS. This situation is unlikely to occur, if a new outcome variable has the same covariates as the one used in optimal selection, or if new covariates are correlated with those used in optimization. The situation is similar to case-control studies where the controls for a specific outcome can be used for other outcomes~\citep{saarela2012secondary,saarela2008nested} although they are not optimal.

The selection procedure presented in this paper requires that up-to-date survival information is already available during the study. The information of the measured covariates is also needed during the study if it is used in the optimal selection. The proposed approach is also applicable to some retrospective designs. For instance, consider a study where blood samples or other biological specimen are collected and stored for all individuals and years later some biomarkers are measured from the stored sample. Selecting only a subcohort for these measurements may be a reasonable option if the extraction of the biomarkers is expensive. Then an approach similar to one presented in this paper could be used to optimally select the subcohort.

The proposed selection method is non-random in the sense that individuals are selected deterministically according to the selection criterion, but we do not see this as a disadvantage when the parameters of a survival model are of interest. Distributions of the covariates or absolute risks of the outcome in the population can also be assessed using the estimated analysis model. If it is important to assess the distributions of the covariates without relying on model assumptions, the subcohort should be selected as a random sample~\citep{morgamcasecohort}. In this case a randomized version of the sequential design construction could be considered~\citep{atkinson2014randomised}. Our method may introduce some selection bias, if the effect of a covariate changes with age and this has not been taken into account in the selection. The procedure could also be developed to include the costs of data collection in the optimization~\citep{fedorov2013optimal}.

\section*{Acknowledgements}

Hanna Tolonen from National Institute for Health and Welfare is acknowledged for providing the data of the East--West study. The authors thank CSC -- IT Center for Science Ltd. for providing computing resources and Marie Reilly, Kari Auranen and Juho Kopra for helpful comments. The research of the first author was supported by the Emil Aaltonen Foundation.

\bibliographystyle{plainnat}
\bibliography{optdesign,surv}

\end{document}